\documentclass[aps,amsfonts,amsmath,prd,preprint,nofootinbib]{revtex4}
\newcommand{\beq}{\begin{equation}}
\newcommand{\eeq}{\end{equation}}

\usepackage{epsfig,bbm,cancel}

\begin{document}

\title{Bubbles from Nothing}

\author{Jose J. Blanco-Pillado, Handhika S. Ramadhan, and Benjamin Shlaer}
\affiliation{Institute of Cosmology, Department of Physics and Astronomy,\\ 
Tufts University, Medford, MA 02155, USA}

\def\changenote#1{\footnote{\bf #1}}

\begin{abstract}
Within the framework of flux compactifications, we construct 
an instanton describing the quantum creation of an open universe from nothing.
The solution has many features in common with the smooth 6d 
bubble of nothing solutions discussed recently, where the 
spacetime is described by a 4d compactification of a 6d 
Einstein-Maxwell theory on $S^2$ stabilized by flux. The 
four-dimensional description of this instanton reduces to 
that of Hawking and Turok.  The choice of parameters uniquely
determines all future evolution, which we additionally find to be stable
against bubble of nothing instabilities.
\end{abstract}

\maketitle
\thispagestyle{empty}
\section{Introduction}
\setcounter{page}{1}

Quantum gravitational effects must be important if the early universe were ever
at temperatures of order the Planck scale.  This would be the case if
slow-roll inflation were preceded by a Big Bang, where the initial
singularity marks where the semi-classical theory has
completely broken down.
Calculations near this regime remain prohibitively difficult. Nevertheless,
it is possible that quantum gravitational effects can be relevant at
far lower energy scales than $T_{\mbox{\tiny Pl}}$. This possibility
is assumed in the field of quantum cosmology,
where the semi-classical universe emerges from non-perturbative
effects  in quantum gravity, but at a scale where
solutions can be trusted.  This is called ``creation of the universe
from nothing"  \cite{Vilenkin-82,Hartle-Hawking-83,Linde-84,Rubakov-84,Vilenkin-84} , 
although it should be thought of as the emergence of a semi-classical
universe from a regime not describable as a classical spacetime. In
principle, the choice of parameters is the only freedom, and initial conditions are
determined solely by  the instantons and their Euclidean action.

The quantum creation of the universe from nothing
has, over the years, attracted a lot of attention. 
Most of the models discussed in the literature studied the formation 
of a closed universe, since it was difficult to see how a spatially 
infinite flat or open universe could emerge from a compact
instanton.\footnote{One can however create a flat or open universe 
by a compact instanton of non-trivial topology, see for example 
 \cite{Zeldovich-Starobinsky}, \cite{Linde-04}.} On
the other hand, the instanton describing the nucleation of a 
bubble universe from a false vacuum provides an example of how to 
go around this objection. The symmetry of the decay process is such 
that the interior of the bubble can be shown to be a spatially
infinite open universe \cite{Coleman-deLuccia,Gott} whose Big Bang 
surface coincides with the lightcone emanating from the nucleation
center of the bubble. Using a somewhat similar strategy, one could 
imagine an instanton that describes the formation of an open 
universe from nothing, in other words, a Euclidean solution 
similar to the bubble nucleation instanton, but without the false vacuum region.
Solutions of this type were first described in
\cite{Hawking-Turok}. One characteristic property of such instantons 
is that they are singular. It is precisely the existence of this 
singularity which eliminates the possibility of including the false 
vacuum region in the Lorentzian continuation of these instantons.
On the other hand, the
singularities are mild enough that their Euclidean action is finite, suggesting
that one should consider these solutions as meaningful contributions
to the path integral. Nevertheless, the validity of these types of
instantons was called into
question by several authors \cite{Linde-98,Bousso-Linde,Vilenkin-98}. In particular, it
was realized in \cite{Vilenkin-98} that the admittance of such singular instantons 
would imply the possibility for Minkowski space to decay by an analogous
process.

The conceptual difficulty introduced by the Hawking-Turok (HT)
singularities motivated people to look for alternative
instantons describing creation of an open universe from nothing. 
One suggestion put forward in \cite{Garriga-1} was to consider the singularity
as resulting from the dimensional
reduction of a smooth higher-dimensional solution. This was
described in \cite{Garriga-1} using a 5d solution
whose $S^1$ extra dimension smoothly degenerates in a
particular region of the four-dimensional space. Looking at this
solution from a purely four-dimensional point of view, one recovers the same
type of singular structure introduced in \cite{Hawking-Turok}. The
higher-dimensional geometry is perfectly smooth, just like a bubble 
of nothing \cite{Witten}. Adding
extra dimensions to the instanton solutions 
forces one to incorporate new fields capable of stabilizing 
the new moduli. In the following we will show that adding these 
new fields has important consequences, not only in the Lorentzian
continuation of the solution but for the Euclidean instanton itself.

An alternative proposal to effectively eliminate the singularities
from Hawking-Turok instantons was presented in 
\cite{Garriga-2, Bousso-Chamblin}. These
solutions are supplemented by the presence of a membrane that 
allows one to cut off the instantons before reaching the
singularity.

\begin{figure}[htbp]
\centering\leavevmode
\epsfysize=8cm \epsfbox{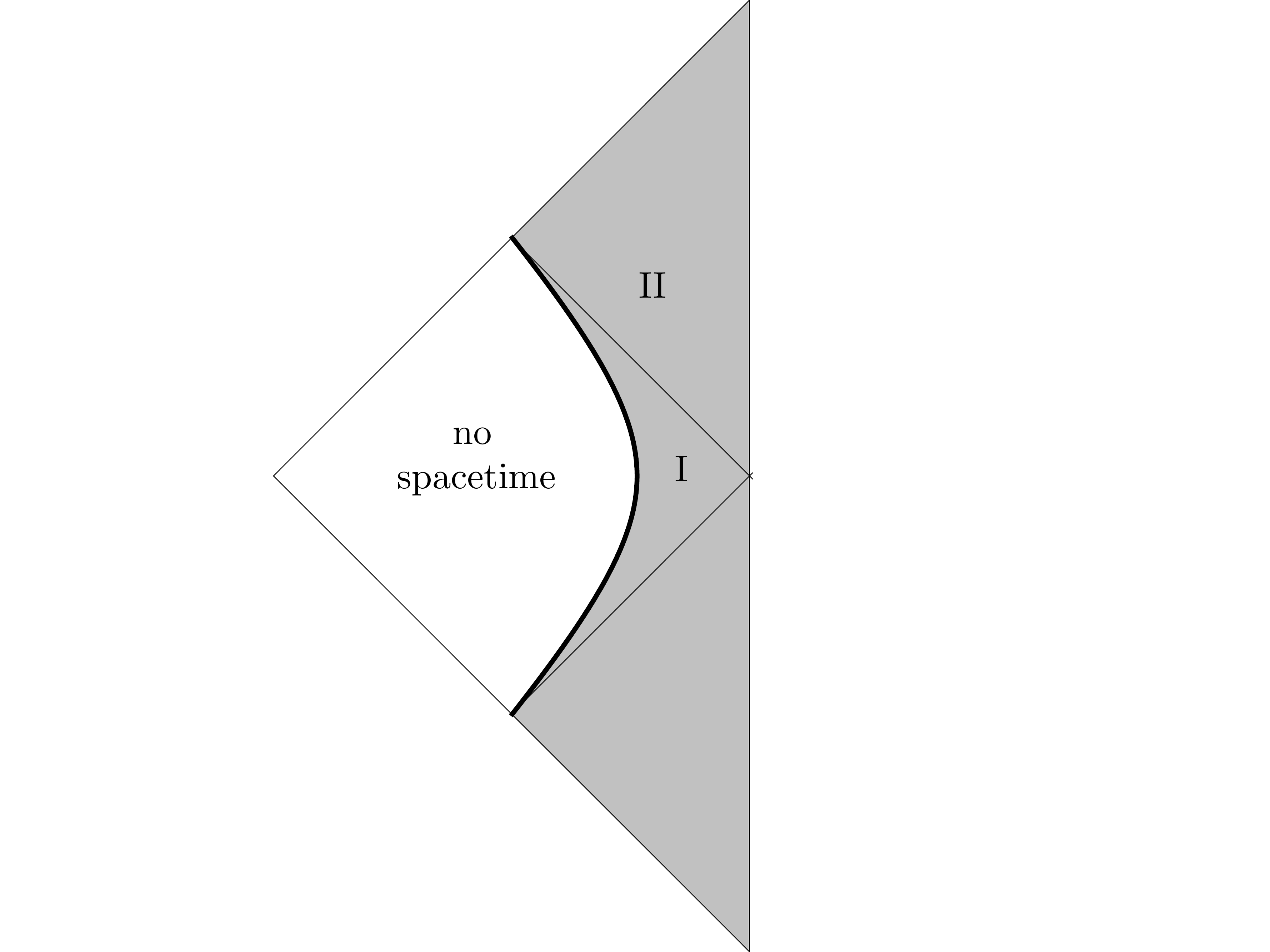}
\caption {The causal structure of a 4d-Minkowski bubble from
  nothing. Only the shaded region exists. From this four-dimensional
  point of view, the domain wall (thick line) is
a singular boundary.  Roughly speaking, the bottom half of this
diagram is virtual, and the nucleation occurs at the narrowest point.}
\label{conformal}
\end{figure}

The instanton solutions we introduce here exhibit characteristics of
both of these resolutions in a natural way.  In our scenario, 
the four-dimensional spacetime is a 6d flux compactification with all
moduli sufficiently stabilized, and
which is manifestly smooth.  The four-dimensional
causal structure is shown in Fig.~(\ref{conformal}).

\begin{figure}[htbp]
\centering\leavevmode
\epsfysize=8cm \epsfbox{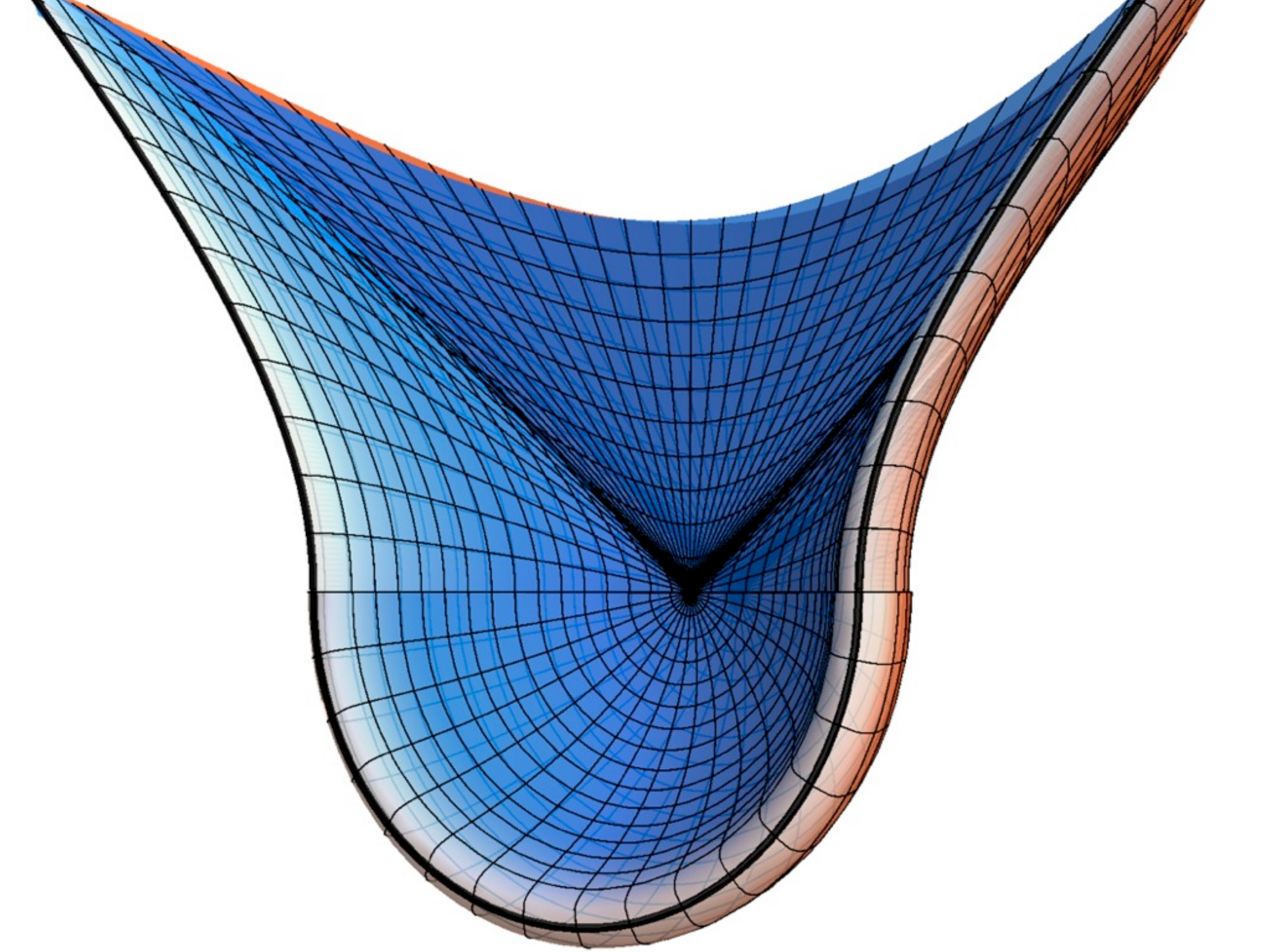}
\caption {An illustration of the 6d smooth Euclidean instanton glued onto the
Lorentzian bubble from nothing.  The core is shown in black surrounded
by a white region indicating unbroken $SU(2)$. The $S^2$ compactification manifold is
represented by $S^0$ (two points), which degenerates at the core.}
\label{3dinstanton}
\end{figure}

We will consider the creation of an open 4d universe from a six
dimensional theory. The existence of extra dimensions allows us to
obtain a compact, smooth solution of the higher-dimensional Euclidean
equations of motion which would be singular from a purely 4d point of
view, similar to what was found in \cite{Garriga-1}. The analytic 
continuation of this solution into the Lorentzian
regime leads to a Kaluza-Klein cosmological scenario. We are therefore 
led to include additional gauge fields in our model whose fluxes
stabilize the extra dimensions and yield a viable cosmology. 
However, adding this flux over the compactified space
creates an obstacle to the smooth degeneration of the extra dimensions. We solve this problem just 
as in the bubble of nothing solutions recently found in flux compactifications \cite{BPS,BPRS}, i.e., by introducing new 
degrees of freedom into the theory that regularize the singularities
with charged solitonic branes at the degeneration loci. One can
think of these branes from
a lower-dimensional perspective in a dualized theory \cite{BPSPV-1} effectively
recovering the idea envisioned in \cite{Garriga-2,Bousso-Chamblin}.  The smooth 6d instanton and its Lorentzian continuation
is illustrated in Fig.~(\ref{3dinstanton}).

The cosmological evolution of flux
compactifications has been extensively studied \cite{KK-cosmology}, including their global
spacetime structure \cite{Linde-transdimensional}, but
observational implications have only recently begun to be 
uncovered \cite{BPSPV-1,CJR,BPSPV-2,SPV,JJ-Salem,Graham,Adamek,Salem,Arnold-et-al}. 
These studies portray a complicated transdimensional multiverse whose {\it pocket bubble
  universes} have different numbers of
large dimensions in an eternally inflating background.
 
Importantly, this intricate multiverse does not admit sensible
solutions eternal to the past \cite{Borde-Guth-Vilenkin}, 
and it is therefore still necessary to seek a theory of initial
conditions. (For related work see \cite{FST}.)

\section{Bubble from nothing universe}

\subsection{Compactification solutions}

Following the arguments given above, we will
study the quantum creation of a 4d open universe in the framework of a 
higher-dimensional theory. A simple scenario which contains enough
richness to build models is a 6d Einstein theory compactified on $S^2$. To
stabilize the $S^2$, we need a two-form field strength, so we include
a Maxwell field. To have smooth magnetically charged
branes in our theory, we
embed this $U(1)$ in $SU(2)$ broken by an adjoint Higgs.  
We must ensure that the extra dimensions are sufficiently stabilized
to remain compactified through subsequent cosmological evolution. 

Here we briefly review the different vacuum solutions found in our
model, described by the 6d action
\begin{equation}
{S}=\int d^{6}x\sqrt{-g}\left(\frac{1}{2\kappa^2}R
-\frac{1}{4}{\cal F}^{a}_{MN}{\cal F}^{a
  MN}-\frac{1}{2}D_{M}\Phi^{a}D^{M}\Phi^{a} - V(\Phi) - \Lambda\right),
\end{equation}
with $M,N = 0,...5$ and $\kappa^2 = M_{\mbox{\scriptsize P}}^{-4}$, where $M_{\mbox{\scriptsize P}}$ is the 6d
reduced Planck mass, and $\Lambda$ denotes the six-dimensional
cosmological constant. The remaining terms in the Lagrangian are
given by
\begin{eqnarray}
V(\Phi)& = &\frac{\lambda}{4}\left(\Phi^{a}\Phi^{a}-\eta^{2}\right)^{2},\\
{\cal F}^{a}_{MN}& = &\partial_{M}A^{a}_{N}-\partial_{N}A^{a}_{M}+ 
e\epsilon^{abc}A^{b}_{M}A^{c}_{N},\\
D_{M}\Phi^{a}& = &\partial_{M}\Phi^{a}+e\epsilon^{abc}A^{b}_{M}\Phi^{c}.
\end{eqnarray}

It was shown long ago \cite{Cremmer} that this theory admits 4d vacua
by turning on a magnetic flux along the internal 2-sphere
compactification manifold. One can easily generalize this
type of solution to arbitrary flux number $n$ using the monopole-like ansatz
\begin{eqnarray}\label{n-hedgehog-metric}
ds^2 &=& {g}_{MN} dx^M dx^N = {g}_{\mu \nu} d x^{\mu} d x^{\nu} 
+ C^2 d\Omega_2^2,\\ \label{n-hedgehog-1}
\Phi^{a}&=&\eta\ p_{c} (\sin\theta \cos n \varphi, \sin\theta\sin n
\varphi, \cos\theta)^a,\\
A^{a}_{\mu}&=&A^{a}_{r}=0,\\
A^{a}_{\theta}&=&\frac{1-w_{c}}{e} (\sin n \varphi,-\cos n \varphi,0)^a,\\
A^{a}_{\varphi}&=&\frac{n\left(1-w_{c}\right)}{e} \sin\theta (\cos\theta \cos n
\varphi, \cos\theta \sin n \varphi, -\sin\theta)^a,\label{n-hedgehog-4}
\end{eqnarray}
with $n \in {\mathbbm{Z}}$. For $n>1$, the equations of motion
constrain the possible values of the constants in the ansatz 
Eqs.~(\ref{n-hedgehog-1}-\ref{n-hedgehog-4}) to be $p_c = 1$
and $w_c = 0$, and the equations of motion reduce to
\begin{eqnarray}
3 H^2 + {1\over {C^2}} &=& \kappa^2 \left({{n^2}\over{2 e^2 C^4}} +
 \Lambda\right), \\
6 H^2  &=&  \kappa^2 \left(\Lambda - {{n^2}\over{2 e^2 C^4}}\right),
\end{eqnarray}
where $H$ denotes the effective 4d Hubble parameter, and $C$
describes the size of the compactification manifold. The generic
solution of these equations is given by \cite{BPRS}
\begin{eqnarray}
C^2 &=& {1\over{ \kappa^2 \Lambda}} \left(1 - \sqrt{1 - {{3 \kappa^4 \Lambda n^2}
\over {2 e^2}}}\right),\\
H^2 &=& {{2 \kappa^2 \Lambda}\over {9 }} \left[1 -  {{ e^2 }\over {3
 \Lambda \kappa^4 n^2}} \left(1 + \sqrt{1 -  {{3 \kappa^4 \Lambda n^2 }\over {2
        e^2}}}\right)\right].
\label{EYMH-landscape-solutions}
\end{eqnarray}
We see from these solutions that the landscape of 4d vacua
includes compactifications of the form $AdS_4 \times S^2$,
${\mathbb R}^{1,3} \times S^2$, and $dS_4 \times S^2$. This is the same
landscape of vacua of Einstein-Maxwell theory in 6d 
\cite{EM6d,BPSPV-1}. The reason for this can be traced back to the
conditions $p=1$ and $w=0$, which effectively
reduce the theory to the abelian Einstein-Maxwell model. Instanton
solutions describing transitions between these vacua have been
recently discussed in several papers
\cite{BPSPV-1,CJR,BPSPV-2,Yang,BD}. Furthermore, in \cite{BPS,
  BPRS}, we identified a new instability for such vacua, the decay 
via a charged bubble of nothing
\cite{Witten}. In this paper we will study, in some sense, the
opposite of the bubble of nothing instability, a process that
creates an open universe from nothing.

\subsection{The instanton solution}

In order to find the gravitational instanton
that interpolates between nothing and one of the compactified
vacuum solutions discussed above, we consider the Euclidean metric
ansatz 
\begin{equation}
\label{eq:Euclidean-metric}
ds^{2}= B^{2}(r)\left(d\psi^{2} + \sin^{2}\psi\,d\Omega_{2}^{2}\right) + dr^{2}
+ C^{2}(r)\left(d\theta^{2} + \sin^{2}\theta\,d\varphi^{2}\right).
\end{equation}
We will show below that the Lorentzian continuation of this 
spacetime will indeed have a region that describes an open FRW universe with
an extra-dimensional $S^2$. Furthermore, the ansatz
for the matter fields is the $r$-dependent generalization of the $n = 1$ flux
compactification solution of
Eqs. (\ref{n-hedgehog-1}-\ref{n-hedgehog-4}) into the Euclidean region, namely,
\begin{eqnarray}
\label{eq:matter}
\Phi^{a}&=& \eta\ p(r) \left(\sin\theta \cos \varphi, \sin\theta
\sin  \varphi, \cos\theta\right)^a,\\
A^{a}_{\mu} &=& A^{a}_{r}=0,\\
A^{a}_{\theta} &=& \frac{1-w(r)}{e} \left(\sin  \varphi,-\cos  \varphi,0\right)^a,\\
A^{a}_{\varphi} &=& \frac{1-w(r)}{e} \sin\theta \left(\cos\theta \cos \varphi, 
\cos\theta \sin  \varphi, -\sin\theta\right)^a.
\end{eqnarray}

There are different boundary conditions that one can
impose on the functions present in this ansatz that would lead to a compact
gravitational instanton. Some of these solutions were already identified in 
the literature, and their interpretations are quite distinct from the one
we present in this paper. Let us briefly mention some of the known
cases.

One could consider an instanton with boundary conditions
\begin{equation}
B(0) = 0, \qquad B(r_h)= 0,\qquad C(0) = C_0,\qquad C(r_h)= C_h.
\end{equation}
Solutions of this type were considered in the purely 
Einstein-Maxwell model where they were interpreted as
the creation of an inflating magnetically charged black brane in
$dS_6$ \cite{BPSPV-2}, or its inverse process, the decompactification transition.
The limiting case where $C_0 = C_h$ is the higher-dimensional analogue of the
Nariai solution.

In this paper we will consider the boundary conditions
\footnote{One could also consider the case
$B(0) = B_0$, $B(r_h)= B_h$, $C(0) = 0$, and $C(r_h)= 0$,
whose Lorentzian continuation could be interpreted as a warped 
$dS_3 \times S^3$ flux compactification, where $S^3$ is the squashed sphere
$dr^{2} + C^{2}(r)(d\theta^{2} + \sin^{2}(\theta)d\varphi^{2})$. We will
not consider such metrics.}
\begin{equation}
\label{instanton-bc}
B(0) = B_0,\quad B(r_h)= 0, \qquad C(0) = 0,\quad C(r_h)= C_h.
\end{equation}
Similar conditions were found in our previous work
describing the decay of $dS_4 \times S^2$ flux vacua via a bubble of nothing. We will now
show that the same boundary conditions can also be used to describe the creation of an open universe from nothing.

In the rest of the paper we will take for simplicity the case $n=1$ and set
\begin{equation}
\Lambda = \frac{e^2}{2 \kappa^4},\nonumber\\
\end{equation}
in other words, we will focus here on finding the instanton asymptotic to a 
Minkowski (${\mathbb R}^{1,3} \times S^2$) compactification, or more precisely 
the open FRW compactification with vanishing effective 4d vacuum energy.

As shown in \cite{BPRS}, the presence of magnetic flux threading the 
extra dimensions can only be made compatible with the boundary
conditions imposed in Eq. (\ref{instanton-bc}) if we incorporate
a solitonic magnetically charged brane at $r=0$.
One can show that the most general smooth solution describing this
core is given by the Taylor expansion
\begin{eqnarray}
p(r) & = & p_{1} r + \cdots,\nonumber\\
w(r) & = & 1 + w_{2}r^{2} + \cdots,\nonumber\\
B(r) & = & B_{0} + B_2 r^{2}+ \cdots,\nonumber\\
C(r) & = & r + C_3 r^{3} + \cdots.
\label{nearcore}
\end{eqnarray}
Note that the local geometry around this point
is similar to the charged generalization of the bubble of nothing
solutions discussed in \cite{BPRS}. Comparing this solution to the
Hawking-Turok instanton, we see that it is this codimension three soliton which
permits a regular geometry, avoiding the singularity present
in the 4d description.

On the other side, near the horizon at $r=r_h$, the solution takes the form
\begin{eqnarray}
\label{dshorizon}
p(r) & = & p_h + p^h_2 (r-r_{h})^{2} + \cdots,\\
w(r) & = & w_h + w^h_2 (r-r_{h})^{2}\cdots,\\
B(r) & = & (r-r_{h}) - B_3^h(r-r_{h})^{3} + \cdots,\\
C(r) & = & C_{h} - C^h_2 (r-r_{h})^{2} +\cdots .
\end{eqnarray}

\begin{figure}[tbp]
\centering\leavevmode
\epsfysize=8cm \epsfbox{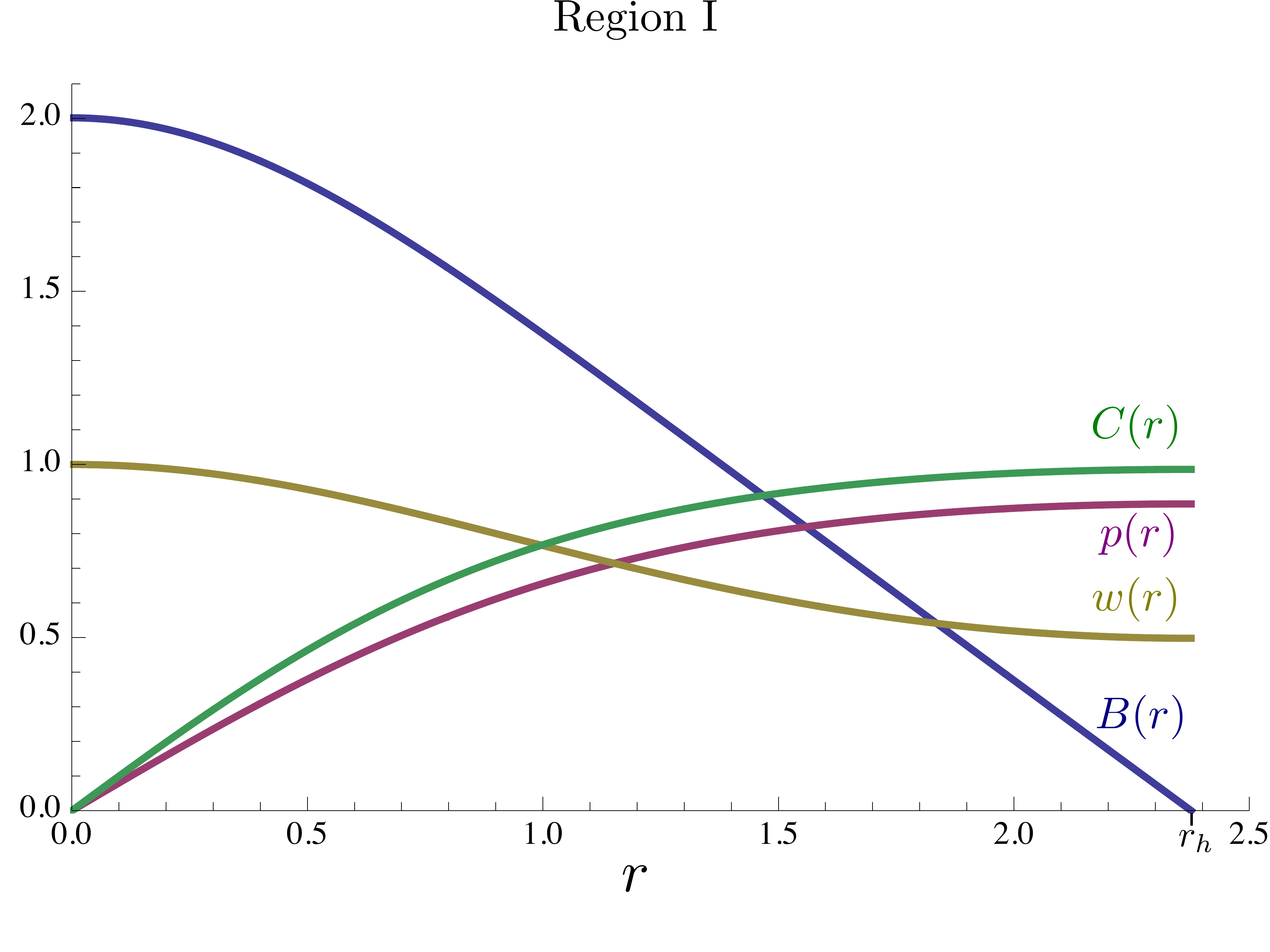}
\caption {The bubble from nothing is characterized by both its Region
  I solution, shown here, and its Region II solution shown below. The
core of the bubble is at $r = 0$, and the horizon at $r = r_h$ leads 
to the future-directed region II of Fig.~(\ref{conformal}). All quantities
are expressed in 6d reduced Planck units (i.e., $\kappa = 1$).}
\label{genregion1}
\end{figure}

\begin{figure}[tbp]
\centering\leavevmode
\epsfysize=12cm \epsfbox{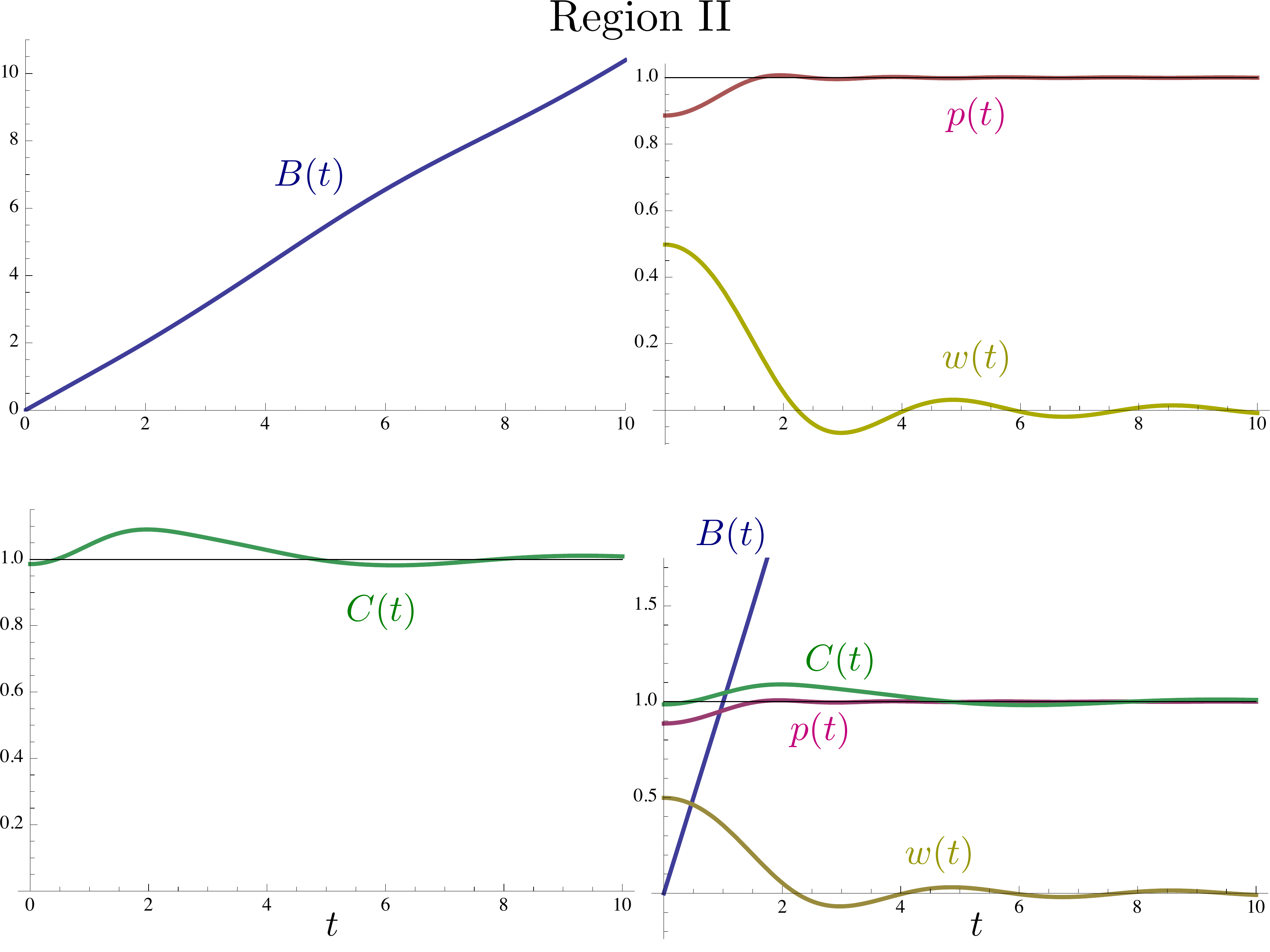}
\caption {The future directed solution of Region II. The scale factor
  $B(t)$ grows linearly due to curvature domination, while the other
fields rapidly approach their vacuum values.}
\label{genallregion2}
\end{figure}

In \cite{BPRS} we described the numerical methods used to 
find explicit solutions which interpolate between these two regions
of the spacetime. The Euclidean solution is identical in form to the
Lorentzian region I solution, an example of which we show in 
Fig.~(\ref{genregion1}) using parameters $\eta = 2$, $\lambda = 3$,
$e=1$, $\Lambda = \frac{1}{2}$. The corresponding
region II is shown in Fig.~(\ref{genallregion2}).

\subsection{The Lorentzian continuation: Open Inflation.}
We can now analytically continue the instanton solution
given in Eq. (\ref{eq:Euclidean-metric}) by taking the $S^3$ polar angle
$\psi \rightarrow {\pi\over 2} + it$, so that we arrive at the Lorentzian metric
\begin{equation}
\label{lorentzian-metric}
ds^{2}= B^{2}(r)\left(-dt^{2} + \cosh^{2}t\,d\Omega_{2}^{2}\right) + dr^{2}
+ C^{2}(r)\left(d\theta^{2} + \sin^{2}\theta\,d\varphi^{2}\right).
\end{equation}
Thus the magnetic soliton at $r = 0$ is inflating, i.e., it is a co-dimension three
de Sitter brane.  A horizon exists at $r = r_h$, where $B(r)$ vanishes
and the extra dimensions have finite size $C_h > 0.$ 
This is quite similar to
what happens in the spacetime geometry of a global string
\cite{Gregory}. The difference is that in our case 
the region beyond the horizon is stably compactified by the magnetic
flux along the $S^2$ extra dimensions. One can obtain a
geometry similar to the one described here in the global string case,
along the lines of \cite{BPS}.

We extend the geometry beyond the horizon to obtain
\begin{equation}
\label{inside-lightcone}
ds^{2}= - dt^2 + B^{2}(t) d{\cal H}^2_3 
+ C^{2}(t) d\Omega^2_2.
\end{equation}
This is the line element of an anisotropic six-dimensional
universe. The interesting point is that one can choose the parameters of
the theory such that the subsequent evolution is in the basin of attraction of a flux vacuum,
i.e., a four-dimensional open universe.

Any viable model should ensure the existence of a period of slow-roll
inflation within the bubble universe.
A realistic treatment is beyond the scope of this paper, and it seems likely one will need to add
new ingredients to our toy model to satisfy observational constraints.  

Nevertheless we have found an interesting point in parameter space 
that demonstrates $\approx 17$ e-folds 
of 4d slow-roll inflation without adding additional fields (or tuning
of initial conditions, which are fixed). This is shown in Figs.~(\ref{infregion1}-\ref{infallregion2}) below. By fine-tuning
the parameters in the Lagrangian, the bubble solution propels
the radion $C(t)$ outward to the top of the effective compactification potential.
This in turn means that the evolution in the interior of the bubble is 
quickly dominated by the 4d effective vacuum energy associated
with this potential, and so the effective 4d scale factor soon
begins to grow exponentially. One can identify three different epochs for the
evolution of the scale factor $B(t)$ in Fig.(\ref{infallregion2}). Like
all bubble interiors, this universe begins curvature dominated, and
$B(t)$ grows linearly. This is rapidly followed by a vacuum energy
dominated epoch which corresponds to a period of
slow-roll inflaton. Finally, the radion suddenly rolls toward its vacuum,
initiating an epoch of matter domination due to the massive
field oscillations about the potential minimum. In a realistic 
setup the inflaton $C(t)$ must couple to the standard model and reheat.

To achieve the inflationary solution, we chose parameters 
\begin{equation}\label{parameters}
\Lambda = \frac{1}{2},\qquad e = 1,\qquad \eta = 2.4070426,\qquad \lambda = 6.7275,
\end{equation}
in 6d reduced Planck units, where $\kappa = 1$.

\begin{figure}[tbp]
\centering\leavevmode
\epsfysize=8cm \epsfbox{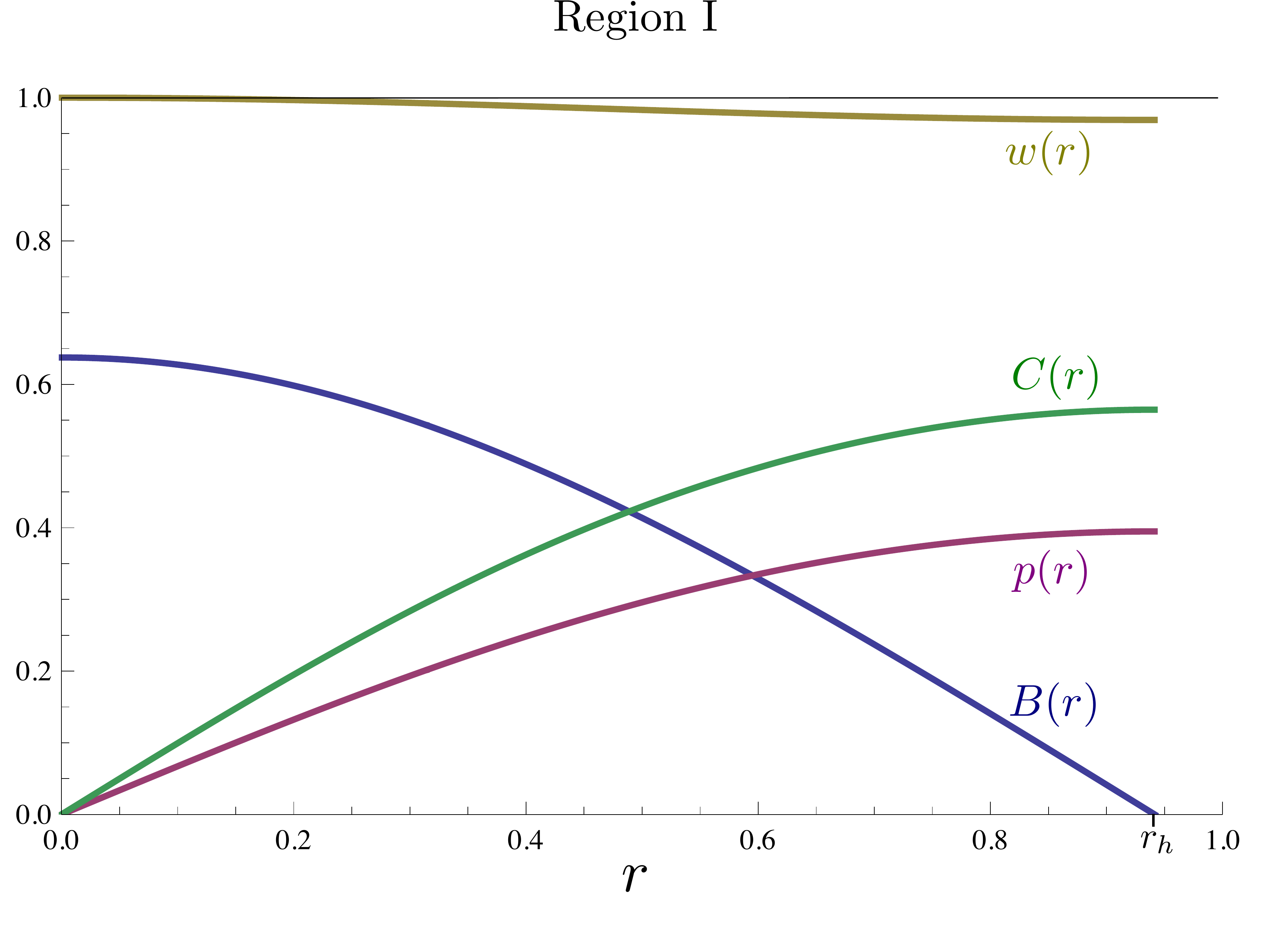}
\caption {Region I of the inflationary solution. The parameters which yield this solution are
$\Lambda = \frac{1}{2}$, $e = 1$, $\eta = 2.4070426$, $\lambda = 6.7275$.}
\label{infregion1}
\end{figure}

\begin{figure}[tbp]
\centering\leavevmode
\epsfysize=12cm \epsfbox{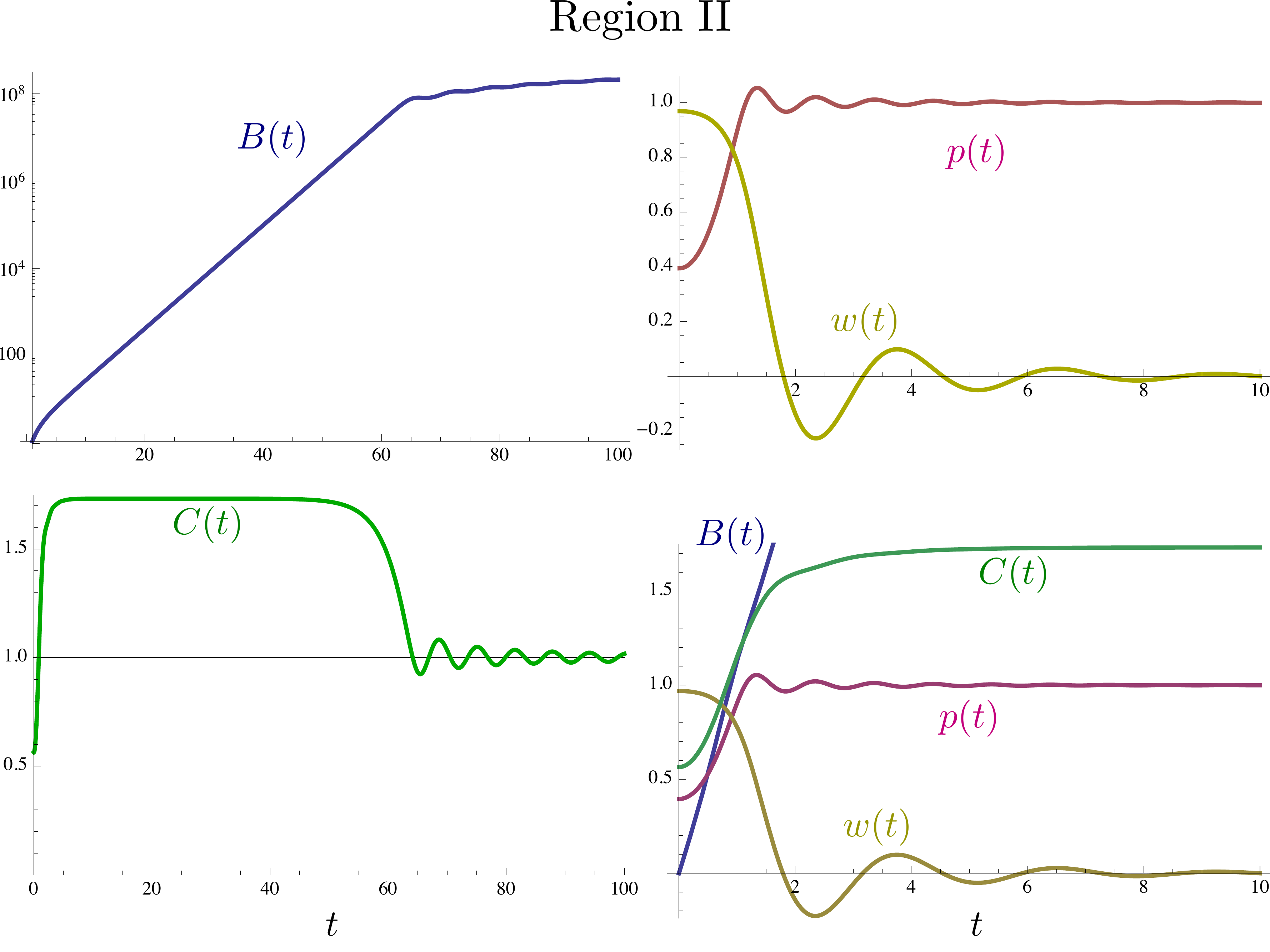}
\caption {The future directed inflationary solution of Region II.  Notice $B(t)$
  grows exponentially before transitioning to power-law growth. While
  the matter fields $p(t)$ and $w(t)$ rapidly approach their stable
  vacuum values, the radion $C(t)$ serves as the inflaton, and only
  falls back to its stable compactification value after $\approx 17$
  e-folds.}
\label{infallregion2}
\end{figure}

\section{Other solutions}

In the previous section we considered parameters resulting in Hawking-Turok
type instantons, where an end-of-the-world brane expands from nothing with
an open universe interior of vanishing 4d cosmological constant.  By adjusting
the ratio of the 6d cosmological constant $\Lambda$ and the gauge coupling $e$,
this can result in $AdS_4$ (big-crunch) or $dS_4$ cosmologies, as
shown in Eq.~(\ref{EYMH-landscape-solutions}). We will
not consider higher $n$ compactifications, which furthermore do not admit $O(3)$
invariant solitonic sources \cite{Weinberg-Guth}.

Within the same vacuum energy sector, 
one can further vary parameters in the theory, namely
the scalar self-coupling $\lambda$ and the symmetry breaking scale $\eta$.
These change the thickness and tension
of the solitonic brane without altering the vacuum energy.   
The full four-dimensional (excluding $n$) parameter space will be
given in a more complete description elsewhere. For now, we content ourselves with the
study of the one-dimensional subspace of solutions with different values of $\eta$, and all
other parameters fixed.  We will see that this is sufficient to probe
a wide variety of effective 4d ``domain wall" tensions.
Note that our model does not not admit stable compactifications 
for $\eta < 1$ in 6d reduced Planck units, so we will not comment further on this range.

\subsection{Bubbles of Nothing. ($1 < \eta < \eta_c$)}
Following \cite{BPS, BPRS}, one can find bubble of nothing solutions
in our model.  These differ by having an apparently negative tension
4d effective domain wall,
since the bubble is gravitationally attractive.  They are interpreted as instantons
contributing to the decay rate of flux compactifications to
nothing.  For small values of $\eta >1$,
one finds a geometry very similar to the generalized
version of the original bubble of nothing \cite{Witten}. From a 6d
point of view they describe a gravitationally attractive
``throat'' where the $S^2$ extra dimensions smoothly degenerate as one
approaches the soliton core. Increasing $\eta$, a small
repulsive region develops near the core
of the object, but long range attraction remains. This is the type of solutions that we
  previously referred to as the ``punted bubble of nothing."  Further increasing of $\eta$ 
 continues to increase the $dS_3$ radius of the bubble surface until the critical value
$\eta_c$, at which point the radius is infinite, i.e., the bubble is flat.\footnote{Incidentally
  this is similar to the kind of behavior that was suggested in
  \cite{Gia}.} Solutions similar to these can be found in
a different context in \cite{Roessl, alexcho}.

\subsection{The Critical Bubble. ($\eta = \eta_c$)}
One can choose the parameters such that the (2+1)d brane worldvolume is a perfectly flat
``domain wall" end to a ${\mathbbm R}^{1,3}$ or $AdS_4$ flux
compactification\footnote{Note that the radius of curvature of a
  domain wall can never exceed the radius of curvature of the
  bulk. Hence the flattest domain wall in $dS_4$ has positive
  curvature, and $AdS_3$ domain walls only exist in
  $AdS_4$.}. Solutions of the $AdS_4$ type resemble smooth higher-dimensional versions \cite{cigar-compactification} of Randall-Sundrum \cite{RSII}
compactifications, albeit to a 3d universe in our case. (See
also \cite{Roessl, alexcho} for related solutions.)  One example
of a flat bubble is given by the parameters  $\eta_{c}=1.447928125$ , $\lambda = 6.7275$, $e = 1$, $\Lambda = 1/2$.

\subsection{Bubbles from Nothing. ($\eta_c < \eta< \eta_d$)}
Further increasing $\eta$ results in another inflating magnetic
brane solution, but with different boundary conditions, with a
horizon at some distance away from the core. This is the type of
solution we described above, which
resembles the domain wall solutions of \cite{Vilenkin-wall,Ipser,Garriga-Sasaki} and global
cosmic string solutions of \cite{Gregory}. As we explained, looking
at these solutions beyond the horizon reveals a stable compactification,
and they are therefore interpreted as instantons describing the
creation of an open 4d universe from nothing. Increasing 
$\eta$ changes the value of the matter fields at the horizon,
effectively changing the details of the 4d universe in this
region. We have seen an example of this in the main part of the text,
where we discussed the effect of the value of $\eta$ on the amount
of inflation.

\subsection{Decompactification. ($\eta > \eta_d$)}

By increasing the value of $\eta$ beyond $\eta_d$,
the open universe fails to remain compactified.
Instead, the radion $C(t)$ rolls to infinity, and the Region II 
geometry is actually 6d de Sitter space.  This solution
should be thought of as simply a smooth magnetically charged brane in $dS_6$. 
The inflationary solution in
Figs.~(\ref{infregion1}-\ref{infallregion2}) falls just below this
threshold, so the point
$\Lambda = \frac{1}{2}$, $e = 1$, $\eta = 2.4070426$, $\lambda = 6.7275$ is
very near the boundary to decompactification.

Finally, in the limit of large $\eta$ one encounters the situation
where the values of the matter fields necessarily do not change 
from those at the core of the defect, i.e., $w_h = 1, p_h = 0$. One can think of these 
solutions as describing a monopole brane whose core is larger than the Hubble 
distance of $dS_6$. This is nothing more than topological inflation
\cite{Vilenkin-ti,Linde-ti} in the 6d theory.  A point in parameter space just within this regime
is $\Lambda = \frac{1}{2}$, $e = 1$, $\eta = 2.57$, $\lambda = 6.7275$.

It is interesting to note that the instanton solutions presented in
this case pick out an anisotropic slicing of $dS_6$ associated with the 
orientation of the defect. One could thus envision a
four-dimensional version of this scenario in which the instanton
singles out an anisotropic slicing of 4d de Sitter space.
We will describe such a case in a subsequent paper.

\section{Conclusions}

We have identified 6d instantons that 
describe the creation from nothing of an open 4d universe in 
the context of flux compactifications. The higher-dimensional 
nature eliminates the singularities found in the 4d instantons of this type
\cite{Hawking-Turok}.
The core of the brane allows for
a smooth degeneration of the extra-dimensional space, acting at the same time
as a source for the magnetic flux that is ultimately responsible for the
stabilization, in the asymptotic region, of the modulus
associated with the extra dimensions.

The region of the geometry near the brane resembles an inside-out version of the bubble
of nothing solutions previously investigated
\cite{BPRS}. In a sense this type of
instanton causes the ``opposite" process as the bubble of nothing decay. 
This reasoning might suggest that a universe created from a
bubble {\it from} nothing could decay by a bubble {\it of} nothing.
We found evidence here that this is not the case, since the parameters of the theory alone seem to
determine the 4d effective tension of the domain wall leading to
nothing. We have found a monotonic deformation of bubble of nothing solutions to
bubble from nothing solutions as one increases the tension of the
brane. The critical solution between these two classes is a flat
brane, which in turn suggests that flux compactifications reach an island of stability
against decay via bubbles of nothing in a similar manner that
Coleman-De Luccia transitions are absent for sufficiently negative vacuum energy \cite{Coleman-deLuccia}.
This also suggests that a supersymmetric version of our
theory will possess critical bubble solutions, rendering it stable with
respect to bubble of nothing decays, as expected on general grounds.
Due to the impossibility of Minkowski domain walls in de Sitter space,  the
Bubble of nothing $\leftrightarrow$ Bubble from nothing transition is
less sharp in de Sitter flux compactifications.

Most of the bubble from nothing solutions of the
type described in this paper would not yield a realistic 4d 
universe, since they do not have
enough slow-roll inflation within the bubble. Nevertheless, we have found a
neighborhood in parameter space that leads to a significant number
of e-foldings.  While the parameters are fine-tuned, the initial conditions are
entirely determined by regularity of the instanton.  Such initial conditions would seem completely fine tuned
without quantum cosmology, since we would have to position the
inflaton at the top of its potential to tremendous accuracy by hand. 

The only dynamical quantity which we chose by hand is the flux number $n$, although this is in principle determined
by the corresponding value of the Euclidean action. 
Interestingly, this opens up the possibility of having a discrete prediction for
values of the curvature parameter $\Omega_k$ today, alleviating the 
problems with the continuum of solutions found in the Hawking-Turok
instantons. This is related to the fact that our instanton
is not singular, so we do not have the extra degree
of freedom found in the original HT instantons. 
Our instanton could also lead to distinctive signatures in cosmological perturbations
that could be investigating along the lines of \cite{HT-perturbations}.

We should mention that even if the 4d universe obtained from our toy model
were to have enough e-foldings of slow-roll inflation, one must additionally ensure that
the spectrum of perturbations is consistent with observations possibly requiring the 
inclusion of further ingredients in the theory to flatten and lower the inflaton potential.

\acknowledgments
J.~J.~B.-P.~and B.~S.~would like to thank the Yukawa Institute for
hospitality and support during the Gravity and Cosmology GC2010
workshop, where part of this work was completed. We would also like to
thank Roberto Emparan, Ben Freivogel, Jaume Garriga, Oriol Pujolas,
Mike Salem and Alex Vilenkin for helpful discussions.
J.~J.~B.-P.~is supported in part by the National Science Foundation
under grant 0969910.  B.~S. received support in part by Foundational 
Questions Institute grant RFP2-08-26A.

\end{document}